\documentclass{llncs}
\usepackage{version}
\usepackage{ma}

\title{Instruction Sequences Expressing \\ Multiplication Algorithms}
\author{J.A. Bergstra \and C.A. Middelburg}
\institute{Informatics Institute, Faculty of Science, University of
           Amsterdam, \\
           Science Park~904, 1098~XH Amsterdam, the Netherlands \\
           \email{J.A.Bergstra@uva.nl,C.A.Middelburg@uva.nl}}

\begin{document}
\maketitle

\begin{abstract}
For each function on bit strings, its restriction to bit strings of any 
given length can be computed by a finite instruction sequence that 
contains only instructions to set and get the content of Boolean 
registers, forward jump instructions, and a termination instruction.
We describe instruction sequences of this kind that compute the function 
on bit strings that models multiplication on natural numbers less than 
$2^N$ with respect to their binary representation by bit strings of 
length $N$, for a fixed but arbitrary $N > 0$, according to the long 
multiplication algorithm and the Karatsuba multiplication algorithm.
We find among other things that the instruction sequence expressing the 
former algorithm is longer than the one expressing the latter algorithm 
only if the length of the bit strings involved is greater than $2^8$.
We also go into the use of an instruction sequence with backward jump 
instructions for expressing the long multiplication algorithm.
This leads to an instruction sequence that it is shorter than the other 
two if the length of the bit strings involved is greater than $2$.
\begin{classcode}
F.1.1, F.2.1.
\end{classcode}
\end{abstract}

\section{Introduction}
\label{sect-intro}

This paper belongs to a line of research in which issues relating to 
various subjects from computer science, including programming language
expressiveness, computability, computational complexity, algorithm 
efficiency, algorithmic equivalence of programs, program verification, 
program performance, program compactness, and program parallelization, 
are rigorously investigated thinking in terms of instruction sequences.
An enumeration of most papers belonging to this line of research is 
available at~\cite{SiteIS}.
The work on computational complexity presented in~\cite{BM13a,BM14e} and 
the work on algorithmic equivalence of programs presented 
in~\cite{BM14a} were prompted by the fact that, for each function on bit 
strings, its restriction to bit strings of any given length can be 
computed by a finite instruction sequence that contains only 
instructions to set and get the content of Boolean registers, forward 
jump instructions, and a termination instruction.

This fact also incited us to look for finite instruction sequences 
containing only the above-mentioned instructions that compute a 
well-known function on bit strings of a given length.
In~\cite{BM13b}, we did so taking the hash function SHA-256 from the 
Secure Hash Standard~\cite{NIST12a} as the well-known function on bit 
strings.
In the current paper, we do so taking the function that models 
multiplication on natural numbers less than $2^N$ with respect to their 
binary representation by bit strings of length $N$, for a fixed but 
arbitrary $N > 0$, as the well-known function on bit strings.

We describe finite instruction sequences containing only the 
above-mentioned instructions that compute this function according to the 
standard multiplication algorithm, which is known as the long 
multiplication algorithm, and according to the Karatsuba multiplication 
algorithm~\cite{Kar95a,KO62a}.
We calculate the exact size of the instruction sequence expressing the 
long multiplication algorithm and lower and upper estimates for the size
of the instruction sequence expressing the Karatsuba multiplication 
algorithm.
We find among other things that the instruction sequence expressing the 
former algorithm is longer than the instruction sequence expressing the 
latter algorithm  only if the length of the bit strings involved is 
greater than $2^8$.

We also go into the use of an instruction sequence with backward jump 
instructions for expressing the long multiplication algorithm.
We describe a finite instruction sequence containing a backward jump 
instruction, in addition to the above-mentioned instructions, that 
expresses a minor variant of the long multiplication algorithm.
We calculate the exact size of this instruction sequence and find among 
other things that it is shorter than the other two if the length of the 
bit strings involved is greater than $2$.
In addition, we argue that the instruction sequences expressing the
long multiplication algorithm form a hard witness of the inevitable 
existence of a halting problem in the practice of impera\-tive
programming.

The Karatsuba multiplication algorithm was devised by Karatsuba in 1962 
to disprove the conjecture made by Kolmogorov that any algorithm to 
compute the function that models multiplication on natural numbers with
respect to their representations in the binary number system has time 
complexity $\Omega(n^2)$.
Shortly afterwards, this divide-and-conquer algorithm was generalized by 
Toom and Cook~\cite{Coo66a,Too63a}.
Later, asymptotically faster multiplication algorithms, based on fast 
Fourier transforms, were devised by Sch\"{o}nhage and 
Strassen~\cite{SS71a} and F\"{u}rer~\cite{Fur09a}.
To our knowledge, except for the Sch\"{o}nhage-Strassen algorithm, only 
informal (natural language or pseudo code) descriptions of these 
multiplication algorithms are available.
In this paper, we provide a mathematically precise alternative to the
informal descriptions of the Karatsuba multiplication algorithm, using
terms from an algebraic theory of single-pass instruction sequences
introduced in~\cite{BL02a}.

It is customary that computing practitioners phrase their explanations 
of issues concerning programs from an empirical perspective such as 
the perspective that a program is in essence an instruction sequence.
An attempt to approach the semantics of programming languages from this 
perspective is made in~\cite{BL02a}.
The groundwork for the approach is an algebraic theory of single-pass
instruction sequences, called program algebra, and an algebraic theory
of mathematical objects that represent the behaviours produced by
instruction sequences under execution, called basic thread algebra.%
\footnote
{In~\cite{BL02a}, basic thread algebra is introduced under the name
 basic polarized process algebra.
}
The line of research referred to at the beginning of this introduction 
originates from the above-mentioned work on an approach to programming 
language semantics.

The general aim of this line of research is to bring instruction 
sequences as a theme in computer science better into the picture.
This is the general aim of the work presented in the current paper as 
well.
However, different from usual in the work referred to above, the accent 
is this time mainly on a practical problem, namely the problem to devise 
instruction sequences that express the long multiplication algorithm and 
the Karatsuba multiplication algorithm.
As in the work referred to above, the work presented in the current 
paper is carried out in the setting of program algebra.

This paper is organized as follows.
First, we survey program algebra and the particular fragment and 
instantiation of it that is used in this paper (Section~\ref{sect-PGA})
and sketch the Karatsuba multiplication algorithm 
(Section~\ref{sect-sketch-KMA}).
Next, we describe how we deal with $n$-bit words by means of Boolean 
registers (Section~\ref{sect-words}) and how we compute the operations 
on $n$-bit words that are used in the long multiplication algorithm 
and/or the Karatsuba multiplication algorithm 
(Section~\ref{sect-opns-words}).\linebreak[2]
Then, we describe and analyze instruction sequences that express these 
algorithms (Section~\ref{sect-KMA}).
After this, we go into the use of an instruction sequence with backward 
jump instructions for expressing the long multiplication algorithm
(Sections~\ref{sect-BJMP}) and relate the findings to the halting 
problem (Section~\ref{sect-HP}). 
Finally, we make some concluding remarks (Section~\ref{sect-concl}).

We rely in this paper on an intuitive understanding of what is an 
algorithm and when an instruction sequence expresses an algorithm.
A rigorous study of these issues and related ones, carried out in the 
same setting as the work presented in this paper, is presented 
in~\cite{BM14a}.

The preliminaries to the work presented in this paper are the same as 
the preliminaries to the work presented in~\cite{BM13b}, which are in
turn a selection from the preliminaries to the work presented 
in~\cite{BM13a}.
For this reason, there is some text overlap with those papers.
The preliminaries concern program algebra.
We only give a brief summary of program algebra.
A comprehensive introduction, including examples, can among other things
be found in~\cite{BM12b}.

This paper consolidates material from~\cite{BM13c,BM13d}.

\section{Program Algebra}
\label{sect-PGA}

In this section, we present a brief outline of \PGA\ (ProGram Algebra) 
and the particular fragment and instantiation of it that is used in 
the remainder of this paper.
A mathematically precise treatment can be found in~\cite{BM13a}.

The starting-point of \PGA\ is the simple and appealing perception
of a sequential program as a single-pass instruction sequence, i.e.\ a
finite or infinite sequence of instructions of which each instruction is
executed at most once and can be dropped after it has been executed or
jumped over.

It is assumed that a fixed but arbitrary set $\BInstr$ of
\emph{basic instructions} has been given.
The intuition is that the execution of a basic instruction may modify a 
state and produces a reply at its completion.
The possible replies are $\False$ and $\True$.
The actual reply is generally state-dependent.
Therefore, successive executions of the same basic instruction may
produce different replies.
The set $\BInstr$ is the basis for the set of instructions that may 
occur in the instruction sequences considered in \PGA.
The elements of the latter set are called \emph{primitive instructions}.
There are five kinds of primitive instructions, which are listed below:
\begin{itemize}
\item
for each $a \in \BInstr$, a \emph{plain basic instruction} $a$;
\item
for each $a \in \BInstr$, a \emph{positive test instruction} $\ptst{a}$;
\item
for each $a \in \BInstr$, a \emph{negative test instruction} $\ntst{a}$;
\item
for each $l \in \Nat$, a \emph{forward jump instruction} $\fjmp{l}$;
\item
a \emph{termination instruction} $\halt$.
\end{itemize}
We write $\PInstr$ for the set of all primitive instructions.

On execution of an instruction sequence, these primitive instructions
have the following effects:
\begin{itemize}
\item
the effect of a positive test instruction $\ptst{a}$ is that basic
instruction $a$ is executed and execution proceeds with the next
primitive instruction if $\True$ is produced and otherwise the next
primitive instruction is skipped and execution proceeds with the
primitive instruction following the skipped one --- if there is no
primitive instruction to proceed with,
inaction occurs;
\item
the effect of a negative test instruction $\ntst{a}$ is the same as
the effect of $\ptst{a}$, but with the role of the value produced
reversed;
\item
the effect of a plain basic instruction $a$ is the same as the effect
of $\ptst{a}$, but execution always proceeds as if $\True$ is produced;
\item
the effect of a forward jump instruction $\fjmp{l}$ is that execution
proceeds with the $l$th next primitive instruction of the instruction
sequence concerned --- if $l$ equals $0$ or there is no primitive
instruction to proceed with, inaction occurs;
\item
the effect of the termination instruction $\halt$ is that execution
terminates.
\end{itemize}

To build terms, \PGA\ has a constant for each primitive instruction and 
two operators. 
These operators are: the binary concatenation operator ${} \conc {}$ and 
the unary repetition operator ${}\rep$.
We use the notation $\Conc{i = 0}{n} P_i$, where $P_0,\ldots,P_n$ are 
\PGA\ terms, for the PGA term $P_0 \conc \ldots \conc P_n$.
We also use the notation $P^n$. 
For each \PGA\ term $P$ and $n > 0$, $P^n$ is the \PGA\ term defined by 
induction on $n$ as follows: $P^1 = P$ and $P^{n+1} = P \conc P^n$.

The instruction sequences that concern us in the remainder of this paper 
are the finite ones, i.e.\ the ones that can be denoted by closed \PGA\ 
terms in which the repetition operator does not occur. 
Moreover, the basic instructions that concern us are instructions to set 
and get the content of Boolean registers.
More precisely, we take the set
\begin{ldispl}
\set{\inbr{i}.\getbr \where i \in \Natpos} \union
\set{\outbr{i}.\setbr{b} \where i \in \Natpos \Land b \in \Bool}
\\ \;\; {} \union
\set{\auxbr{i}.\getbr \where i \in \Natpos} \union
\set{\auxbr{i}.\setbr{b} \where i \in \Natpos \Land b \in \Bool} 
\end{ldispl}%
as the set $\BInstr$ of basic instructions.

Each basic instruction consists of two parts separated by a dot.
The part on the left-hand side of the dot plays the role of the name of 
a Boolean register and the part on the right-hand side of the dot plays 
the role of a command to be carried out on the named Boolean register.
For each $i \in \Natpos$:
\begin{itemize}
\item
$\inbr{i}$ serves as the name of the Boolean register that is used as 
$i$th input register in instruction sequences;
\item
$\outbr{i}$ serves as the name of the Boolean register that is used as
$i$th output register in instruction sequences;
\item
$\auxbr{i}$ serves as the name of the Boolean register that is used as 
$i$th auxiliary register in instruction sequences.
\end{itemize}
On execution of a basic instruction, the commands have the following 
effects:
\begin{itemize}
\item
the effect of $\getbr$ is that nothing changes and the reply is the 
content of the named Boolean register;
\item
the effect of $\setbr{\False}$ is that the content of the named Boolean 
register becomes $\False$ and the reply is $\False$;
\item
the effect of $\setbr{\True}$ is that the content of the named Boolean 
register becomes $\True$ and the reply is $\True$.
\end{itemize}

Let $n,m \in \Nat$, let $\funct{f}{\set{0,1}^n}{\set{0,1}^m}$, and let
$X$ be a finite instruction sequence that can be denoted by a closed 
\PGA\ term in the case that $\BInstr$ is taken as specified above.
Then $X$ \emph{computes} $f$ if there exists a $k \in \Nat$ such that 
for all $b_1,\ldots,b_n \in \Bool$: if $X$ is executed in an environment 
with $n$ input registers, $m$ output registers, and $k$ auxiliary 
registers, the content of the input registers with names 
$\inbr{1},\ldots,\inbr{n}$ are $b_1,\ldots,b_n$ when execution starts, 
and the content of the output registers with names 
$\outbr{1},\ldots,\outbr{m}$ are $b'_1,\ldots,b'_m$ when execution 
terminates, then $f(b_1,\ldots,b_n) = b'_1,\ldots,b'_m$.

\section{Sketch of the Karatsuba Multiplication Algorithm}
\label{sect-sketch-KMA}

Suppose that $x$ and $y$ are two natural numbers with a binary 
representation of $n$ bits. 
As a first step toward multiplying $x$ and $y$,
split each of these representations into a left part of length 
$\floor{n/2}$ and a right part of length $\ceil{n/2}$.
Let us say that the left and right part of the representation of $x$ 
represent natural numbers $x_L$ and $x_R$ and the left and right part of 
the representation of $y$ represent natural numbers $y_L$ and $y_R$.
It is obvious that $x = 2^\ceil{n/2} \mul x_L + x_R$ and 
$y = 2^\ceil{n/2} \mul y_L + y_R$.
From this it follows immediately that
\begin{ldispl} 
x\mul y = 
2^{2 \mul \ceil{n/2}} \mul (x_L \mul y_L) + 
2^\ceil{n/2} \mul (x_L \mul y_R + x_R \mul y_L) + x_R \mul y_R\;.
\end{ldispl}%
In addition to this, it is known that
\begin{ldispl} 
x_L \mul y_R + x_R \mul y_L = 
(x_L + x_R) \mul (y_L + y_R) - x_L \mul y_L - x_R \mul y_R\;.
\end{ldispl}%
Moreover, it is easy to see that multiplications by powers of $2$ are 
merely bit shifts on the binary representation of the natural numbers 
involved.
All this means that, on the binary representations of $x$ and $y$, the 
multiplication $x \mul y$ can be replaced by three multiplications: 
$x_L \mul y_L$, $x_R \mul y_R$, and $(x_L + x_R) \mul (y_L + y_R)$.
These three multiplications concern natural numbers with binary 
representations of length $\floor{n/2}$, $\ceil{n/2}$, and 
$\ceil{n/2} + 1$, respectively.
For each of these multiplications it holds that, if the binary 
representation length concerned is greater than $3$, the multiplication 
can be replaced by three multiplications of natural numbers with binary 
representations of even shorter length. 

The \emph{Karatsuba multiplication algorithm} is the algorithm that 
computes the binary representation of the product of two natural numbers 
with binary representations of the same length by dividing the 
computation into the computation of the binary representations of three 
products as indicated above and doing so recursively until it not any 
more leads to further length  reduction. 
The remaining products are usually computed according to the standard 
multiplication algorithm, which is known as the long multiplication 
algorithm.

Both the Karatsuba multiplication algorithm and the long multiplication 
algorithm can actually be applied to natural numbers represented in the 
binary number system as well as natural numbers represented in the 
decimal number system.
The long multiplication algorithm is the multiplication algorithm that 
is taught in schools for computing the product of natural numbers 
represented in the decimal number system.
It is known that the long multiplication algorithm has uniform time 
complexity $\Theta(n^2)$ and the the Karatsuba multiplication algorithm 
has uniform time complexity 
$\Theta(n^{\log_2(3)}) = \Theta(n^{1,5849\ldots})$, 
so the Karatsuba multiplication algorithm is asymptotically faster than 
the long multiplication algorithm.

\section{Dealing with $n$-Bit Words}
\label{sect-words}

This section is concerned with dealing with bit strings of length $n$ 
by means of Boolean registers.
It contains definitions which facilitate the description of instruction 
sequences that express the long multiplication algorithm and the 
Karatsuba multiplication algorithm.

Henceforth, it is assumed that a fixed but arbitrary positive natural 
number $N$ has been given.
The above-mentioned algorithms compute the binary representation of the 
product of two natural numbers represented by bit strings of the same 
length.
In Section~\ref{sect-KMA}, the instruction sequences expressing these 
algorithms will be described for the case where this length is $N$.

In the sequel, bit strings of length $n$ will mostly be called 
\emph{$n$-bit words}.
The prefix ``$n$-bit'' is left out if $n$ is irrelevant or clear from
the context.

Let $\kappa{:}i$ 
($\kappa \in \set{\mathsf{in},\mathsf{out},\mathsf{aux}}$, 
 $i \in \Natpos$) be the name of a Boolean register.
Then $\kappa$ and $i$ are called the \emph{kind} and \emph{number} of 
the Boolean register.
Successive Boolean registers are Boolean registers of the same kind with
successive numbers.
Words are stored by means of Boolean registers such that the successive 
bits of a stored word are the contents of successive Boolean registers.

Henceforth, the name of a Boolean register will mostly be used to refer 
to the Boolean register in which the least significant bit of a word is 
stored.
Let $\kappa{:}i$ and $\kappa'{:}i'$ be the names of Boolean registers 
and let $n \in \Natpos$.
Then we say that $\kappa{:}i$ and $\kappa'{:}i'$ \emph{lead to partially 
coinciding $n$-bit words} if $k = k'$ and $0 < |i - i'| < n$.

The $N$-bit words representing the two natural numbers for which the 
binary representation of their product is to be computed are stored in 
advance of the computation in input registers, starting with the input 
register with number $1$.
It is convenient to have available the names $I_1$ and $I_2$ for the 
input registers in which the least significant bit of these words are 
stored.
The $2N$-bit word representing the product is stored just before the end 
of the computation in output registers, starting with the output 
register with number $1$. 
It is convenient to have available the name $O$ for the output register 
in which the least significant bit of this word is stored.
The words representing intermediate values that arise during the 
computation are temporarily stored in auxiliary registers, starting with 
the auxiliary register with number $1$.

In the case of the Karatsuba algorithm, the binary representation of the 
product of two natural numbers with binary representations of the same 
length is computed by dividing the computation into the computation of 
the binary representations of three products and doing so recursively 
until it not any more leads to further length reduction. 
Therefore, it is convenient to have available, for sufficiently many 
natural numbers $i$, the names $I_1^i$, $I_2^i$ and $O^i$ for the 
auxiliary registers in which the least significant bit of the binary
representations of smaller natural numbers and their product are stored. 
Because at each level of recursion, except the last level, the 
computation of the binary representation of a product involves the 
computation of the binary representations of three products at the next 
level, it is convenient to have available, for sufficiently many natural 
numbers $i$, the names $P_1^i$, $P_2^i$ and $P_3^i$ for the auxiliary 
registers in which the least significant bit of these binary 
representations of products are stored. 

It is also convenient to have available the names $S_1,S_2,T_1,T_2$ for 
the auxiliary registers in which the least significant bit of the words 
that represent the intermediate values that arise, other than the ones 
mentioned in the previous paragraph, are stored.
Moreover, it is convenient to have available the name $c$ for the 
auxiliary register that contains the carry/borrow bit that is repeatedly 
stored when computing the operations that model addition and subtraction 
on natural numbers with respect to their binary representation.

Therefore, we define: 
\begin{ldispl}
\begin{asceqns}
I_1           & \deq & \inbr{1}, \\
I_2 \hsp{.25} & \deq & \inbr{k}
              & \mathrm{where}\; k = N + 1, \\ 
O             & \deq & \outbr{1}, \\
c             & \deq & \auxbr{1}, & \hsp{24.6} \\ 
S_1           & \deq & \auxbr{2}, 
\end{asceqns}
\end{ldispl}%
\begin{ldispl}
\begin{asceqns} 
S_2        & \deq & \auxbr{k}
           & \mathrm{where}\; k = 2 \mul N + 2, \\   
T_1        & \deq & \auxbr{k}
           & \mathrm{where}\; k = 4 \mul N + 2, \\
T_2        & \deq & \auxbr{k}
           & \mathrm{where}\; k = 6 \mul N + 2, \\ 
I_1^i      & \deq & \auxbr{k}
           & \mathrm{where}\; k = 10 \mul N \mul i +  8 \mul N + 2
 & (0 \leq i \leq \ceil{\log_2(N-2)}), \\
I_2^i      & \deq & \auxbr{k}
           & \mathrm{where}\; k = 10 \mul N \mul i +  9 \mul N + 2
 & (0 \leq i \leq \ceil{\log_2(N-2)}), \\
O^i        & \deq & \auxbr{k}
           & \mathrm{where}\; k = 10 \mul N \mul i + 10 \mul N + 2
 & (0 \leq i \leq \ceil{\log_2(N-2)}), \\
P_1^i      & \deq & \auxbr{k}
           & \mathrm{where}\; k = 10 \mul N \mul i + 12 \mul N + 2
 & (0 \leq i \leq \ceil{\log_2(N-2)}), \\
P_2^i      & \deq & \auxbr{k}
           & \mathrm{where}\; k = 10 \mul N \mul i + 14 \mul N + 2
 & (0 \leq i \leq \ceil{\log_2(N-2)}), \\
P_3^i      & \deq & \auxbr{k}
           & \mathrm{where}\; k = 10 \mul N \mul i + 16 \mul N + 2
 & (0 \leq i \leq \ceil{\log_2(N-2)}).
\end{asceqns}
\end{ldispl}%
Here $i$ ranges over natural numbers in the interval with lower endpoint 
$0$ and upper endpoint $\ceil{\log_2(N-2)}$.
This needs some explanation.
\begin{proposition}
\label{prop-recursion-depth}
The recursion depth of the Karatsuba multiplication algorithm applied to 
bit strings of length $N$ is $\ceil{\log_2(N-2)}$.
\end{proposition}
\begin{proof}
Let $n \leq N$.
In the Karatsuba multiplication algorithm, the binary representation of 
the product of two natural numbers with binary representations of length 
$n$ is computed by dividing the computation into the computation of the 
binary representation of a product of two natural numbers with binary 
representations of length $\floor{n/2}$, the binary representation of a 
product of two natural numbers with binary representations of length 
$\ceil{n/2}$, and the binary representation of a product of two natural 
numbers with binary representations of length $\ceil{n/2} + 1$.
The function $f$ defined by $f(n) \deq \ceil{n/2} + 1$ has the following
properties: (a)~$f(n) < n$ iff $n > 3$; and (b)~for $n > 3$, the least 
$m$ such that $f^m(n) = 3$ is $\ceil{\log_2(n-2)}$.
This implies that the recursion depth is $\ceil{\log_2(N-2)}$.
\qed
\end{proof}
Proposition~\ref{prop-recursion-depth} tells us that the maximum level 
of recursion that can be reached is $\ceil{\log_2(N-2)}$.
So there are $\ceil{\log_2(N-2)} + 1$ possible levels of recursion, 
\linebreak[2] viz.\ $0$, \ldots, $\ceil{\log_2(N-2)}$.
This means that there are sufficiently many natural numbers $i$ for 
which the names $I_1^i$, $I_2^i$, $O^i$, $P_1^i$, $P_2^i$, and $P_3^i$ 
have been introduced above.
In Section~\ref{sect-KMA}, we will use the names $I_1^i$, $I_2^i$, 
$O^i$, $P_1^i$, $P_2^i$, and $P_3^i$ at the level of recursion 
$\ceil{\log_2(N-2)} - i$.

\section{Computing Operations on $n$-Bit Words}
\label{sect-opns-words}

This section is concerned with computing operations on bit strings of 
length $n$.
It contains definitions which facilitate the description of instruction 
sequences that express the long multiplication algorithm and the 
Karatsuba multiplication algorithm.

In this section, we will write $\beta \beta'$, where $\beta$ and 
$\beta'$ are bit strings, for the concatenation of $\beta$ and $\beta'$.
In other words, we will use juxtaposition for concatenation.
Moreover, we will use the bit string notation $b^n$.
For $n > 0$, the bit string $b^n$, where $b \in \set{0,1}$, is defined 
by induction on $n$ as follows: $b^1 = b$ and $b^{n+1} = b\,b^n$.

The basic operations on words that are relevant to the long 
multiplication algorithm and/or the Karatsuba multiplication algorithm 
are the operations that model addition, subtraction, and multiplication 
by $2^m$, modulo $2^n$, on natural numbers less than $2^n$, with respect 
to their binary representation by $n$-bit words ($0 < n \leq N$, 
$0 < m < n$).
The operation modeling multiplication by $2^m$ is commonly known as 
``shift left by $m$ positions''.
For these operations, we define parameterized instruction sequences 
computing them in case the parameters are properly instantiated (see 
below):
\begin{ldispl}
\ADD{n}(\srcbri{1}{k_1},\srcbri{2}{k_2},\dstbr{l}) \deq {} 
\\ \quad
c.\setbr{\False} \conc {}
\\ \quad
\Conc{i = 0}{n-1} 
 (\ptst{\srcbri{1}{k_1{+}i}.\getbr} \conc \fjmp{8}   \conc 
  \ptst{\srcbri{2}{k_2{+}i}.\getbr} \conc \fjmp{8}   \conc 
  \ntst{c.\getbr}                   \conc \fjmp{14}  \conc {} 
\\ \quad \phantom{\Conc{i = 0}{n-1} (}
  \dstbr{l{+}i}.\setbr{\True} \conc c.\setbr{\False} \conc
   \fjmp{13} \conc 
  \ptst{\srcbri{2}{k_2{+}i}.\getbr} \conc \fjmp{4}   \conc 
  \ptst{c.\getbr}    \conc \fjmp{7} \conc \fjmp{7}   \conc {} 
\\ \quad \phantom{\Conc{i = 0}{n-1} (} 
  \ptst{c.\getbr}                   \conc \fjmp{5}   \conc 
  \dstbr{l{+}i}.\setbr{\False} \conc c.\setbr{\True} \conc
   \fjmp{3} \conc
  \ptst{\dstbr{l{+}i}.\setbr{\False}} \conc
  \dstbr{l{+}i}.\setbr{\True})\;,
\eqnsep
\SUB{n}(\srcbri{1}{k_1},\srcbri{2}{k_2},\dstbr{l}) \deq {} 
\\ \quad
c.\setbr{\False} \conc {}
\\ \quad
\Conc{i = 0}{n-1} 
 (\ntst{\srcbri{1}{k_1{+}i}.\getbr} \conc \fjmp{8}    \conc 
  \ptst{\srcbri{2}{k_2{+}i}.\getbr} \conc \fjmp{8}    \conc 
  \ntst{c.\getbr}                   \conc \fjmp{14}   \conc {} 
\\ \quad \phantom{\Conc{i = 0}{n-1} (}
  \dstbr{l{+}i}.\setbr{\False} \conc c.\setbr{\False} \conc
   \fjmp{13} \conc 
  \ptst{\srcbri{2}{k_2{+}i}.\getbr} \conc \fjmp{4}    \conc 
  \ptst{c.\getbr}    \conc \fjmp{7} \conc \fjmp{7}    \conc {} 
\\ \quad \phantom{\Conc{i = 0}{n-1} (} 
  \ptst{c.\getbr}                   \conc \fjmp{5}    \conc 
  \dstbr{l{+}i}.\setbr{\True} \conc c.\setbr{\True}   \conc
   \fjmp{3} \conc
  \ntst{\dstbr{l{+}i}.\setbr{\True}} \conc
  \dstbr{l{+}i}.\setbr{\False})\;,
\eqnsep
\SHL{n}{m}(\srcbr{k},\dstbr{l}) \deq {}
\\ \quad
\Conc{i = 0}{n-1-m} 
 (\ptst{\srcbr{k{+}n{-}1{-}m{-}i}.\getbr} \conc 
  \ntst{\dstbr{l{+}n{-}1{-}i}.\setbr{\True}} \conc
  \dstbr{l{+}n{-}1{-}i}.\setbr{\False}) \conc {}
\\[.5ex] \quad
\Conc{i = 0}{m-1} (\dstbr{l{+}m{-}1{-}i}.\setbr{\False})\;,
\end{ldispl}%
where 
$s,s_1,s_2$ range over $\set{\mathsf{in},\mathsf{aux}}$, 
$d$ ranges over $\set{\mathsf{aux},\mathsf{out}}$, and
$k,k_1,k_2,l$ range over $\Natpos$.
For each of these parameterized instruction sequences, all but the last 
parameter correspond to the operands of the operation concerned and the 
last parameter corresponds to the result of the operation concerned.
The intended operations are computed provided that the instantiation of 
the last parameter and the instantiation of none of the other parameters 
lead to partially coinciding $n$-bit words.
In this paper, this condition will always be satisfied.

In the case of addition and subtraction, the intended operation is 
computed according to the long addition algorithm and the long 
subtraction algorithm, respectively.
There are many instruction sequences expressing these algorithms.
The ones defined above are at present the shortest ones that we could 
devise.

\begin{proposition}
\label{prop-basic-operations-correct}
Let $n,m \in \Nat$ be such that $0 < n \leq N$ and $0 < m < n$.
Then the function on bit strings of length $n$ computed by 
\begin{enumerate}
\item
$\ADD{n}(I_1,I_2,O) \conc \halt$ models addition modulo $2^n$ on 
natural numbers less than $2^n$ with respect to their binary 
representation by $n$-bit words;
\item
$\SUB{n}(I_1,I_2,O) \conc \halt$ models subtraction modulo $2^n$ on 
natural numbers less than $2^n$ with respect to their binary 
representation by $n$-bit words;
\item
$\SHL{n}{m}(I_1,O) \conc \halt$ models multiplication by $2^m$ modulo 
$2^n$ on natural numbers less than $2^n$ with respect to their binary 
representation by $n$-bit words.
\end{enumerate}
\end{proposition}
\begin{proof}
In the case of the first and second property, we prove a stronger 
property that also covers the final content of the auxiliary register 
containing the carry/borrow bit.
Each of the stronger properties is easy to prove by induction on $n$ 
with case distinction on the contents of the input registers containing 
the most significant bits of the operands of the operation concerned and 
the content of the auxiliary register containing the carry/borrow bit in 
both the basis step and the inductive step.
The third property is easy to prove by induction on $n$ with case 
distinction on the content of the input register containing the most 
significant bit of the operand of the operation concerned in both the 
basis step and the inductive step.
\qed
\end{proof}

Transferring $n$-bit words ($0 < n \leq N$) is also relevant to the 
multiplication algorithms.
For this, we define parameterized instruction sequences as well.
By one the successive bits in a constant $n$-bit word become the content 
of $n$ successive Boolean registers and by the other the successive bits 
in a $n$-bit word that are the content of $n$ successive Boolean 
registers become the content of $n$ other successive Boolean registers:
\begin{ldispl}
\SET{n}(b_0 \ldots b_{n-1},\dstbr{l}) \deq 
\Conc{i = 0}{n-1} (\dstbr{l{+}i}.\setbr{b_i})\;,
\eqnsep
\MOV{n}(\srcbr{k},\dstbr{l}) \deq 
\Conc{i = 0}{n-1} 
 (\ptst{\srcbr{k{+}i}.\getbr} \conc 
  \ntst{\dstbr{l{+}i}.\setbr{\True}} \conc 
  \dstbr{l{+}i}.\setbr{\False})\;,
\end{ldispl}%
\sloppy
where 
$b_0,\ldots,b_{n-1}$ range over $\set{\False,\True}$,
$s$ ranges over $\set{\mathsf{in},\mathsf{aux}}$, 
$d$ ranges over $\set{\mathsf{aux},\mathsf{out}}$, and
$k,l$ range over $\Natpos$.
In the case of $\MOV{n}$, the intended transfer is performed provided 
that the instantiation of the last parameter and the instantiation of 
the first parameter do not lead to partially coinciding $n$-bit words.
In this paper, this condition will always be satisfied.

\begin{proposition}
\label{prop-transfer-operations-correct}
Let $n \in \Nat$ be such that $0 < n \leq N$.
Then the function on bit strings of length $n$ computed by 
\begin{enumerate}
\item
$\SET{n}(b_0 \ldots b_{n-1},O) \conc \halt$ models the natural number 
constant with binary representation $b_0 \ldots b_{n-1}$;
\item
$\MOV{n}(I_1,O) \conc \halt$ models the identity function on natural 
numbers less than $2^n$ with respect to their binary representation by 
$n$-bit words.
\end{enumerate}
\end{proposition}
\begin{proof}
Each of these properties is trivial to prove by induction on $n$ with 
case distinction on $b_{n-1}$ and the content of the input register 
containing the most significant bits of the operand of the operation, 
respectively, in both the basis step and the inductive step.
\qed
\end{proof}

For convenience's sake, we define some special cases of the 
parameterized instruction sequences for transferring $n$-bit words 
($0 < m < n$):
\begin{ldispl}
\ZPAD{n}{m}(\dstbr{l}) \deq
\SET{n-m}(0^{n-m},\dstbr{l{+}m})\;,
\eqnsep
\MVH{n}{m}(\srcbr{k},\dstbr{l}) \deq
\MOV{m}(\srcbr{k{+}(n{-}m)},\dstbr{l})\;,
\eqnsep
\MVL{n}{m}(\srcbr{k},\dstbr{l}) \deq
\MOV{m}(\srcbr{k},\dstbr{l})\;,
\end{ldispl}%
where 
$s$ ranges over $\set{\mathsf{in},\mathsf{aux}}$, 
$d$ ranges over $\set{\mathsf{aux},\mathsf{out}}$, and
$k,l$ range over $\Natpos$.
$\ZPAD{n}{m}$ is meant for turning a stored $m$-bit word into a stored 
$n$-bit word by zero padding.
$\MVH{n}{m}$ and $\MVL{n}{m}$ are meant for transferring only the $m$ 
most significant bits and the $m$ least significant bits, respectively, 
of a stored $n$-bit word. 

Because $\ceil{n/2} + 1 < n$ iff $n > 3$, the Karatsuba multiplication 
algorithm cannot be used for modeling multiplication on natural numbers 
less than $2^n$ with respect to their binary representation by $n$-bit 
words if $n \leq 3$.
Therefore, we also define a parameterized instruction sequence, in terms 
of the above-mentioned basic operations, that computes the operation
modeling multiplication according to the long multiplication algorithm:
\begin{ldispl}
\MUL{n}(\srcbri{1}{k_1},\srcbri{2}{k_2},\dstbr{l}) \deq 
\\ \quad
\MOV{n}(\srcbri{1}{k_1},S_1) \conc \ZPAD{2n}{n}(S_1) \conc
\SET{2n}(0^{2n},S_2) \conc {}
\\ \quad
\Conc{i = 0}{n-1} 
 (\ntst{\srcbri{2}{k_2{+}i}.\getbr} \conc \fjmp{l_i} \conc   
  \ADD{n+i+1}(S_1,S_2,S_2) \conc \SHL{n+i+1}{1}(S_1,S_1)) \conc {}
\\ \quad
\MOV{2n}(S_2,\dstbr{l})\;,
\\
\mbox{where $l_i = \len(\ADD{n+i+1}(S_1,S_2,S_2)) + 1$}\;,
\end{ldispl}%
where 
$s_1,s_2$ range over $\set{\mathsf{in},\mathsf{aux}}$, 
$d$ ranges over $\set{\mathsf{aux},\mathsf{out}}$, and
$k_1,k_2,l$ range over $\Natpos$.
The additions are done on the fly and the shifts are restricted to 
shifts by one position by shifting the result of all preceding shifts.

\begin{proposition}
\label{prop-MUL-correct}
Let $n \in \Nat$ be such that $0 < n \leq N$.
Then the function on bit strings of length $n$ computed by 
$\MUL{n}(I_1,I_2,O) \conc \halt$ models multiplication on natural 
numbers less than $2^n$ with respect to their binary representation by 
$n$-bit words.
\end{proposition}
\begin{proof}
We prove a stronger property that also covers the final contents of the 
$2n$ successive auxiliary registers starting with the one named $S_1$ 
and the $2n$ successive auxiliary registers starting with the one named 
$S_2$.
This stronger property is easy to prove, using 
Propositions~\ref{prop-basic-operations-correct} 
and~\ref{prop-transfer-operations-correct}, 
by induction on $n$ with case distinction on the content of the input 
register containing the most significant bit of the second operand of 
the operation concerned in both the basis step and the inductive step.
\qed
\end{proof}

The calculation of the lengths of the parameterized instruction 
sequences defined above is a matter of simple additions and 
multiplications. 
The lengths of these instruction sequences are as follows: 
\begin{ldispl}
\len(\SHL{n}{m}(\srcbr{k},\dstbr{l})) = 3 \mul n - 2 \mul m\;, \\
\len(\ADD{n}(\srcbri{1}{k_1},\srcbri{2}{k_2},\dstbr{l})) = 
 21 \mul n + 1\;, \\ 
\len(\SUB{n}(\srcbri{1}{k_1},\srcbri{2}{k_2},\dstbr{l})) =
 21 \mul n + 1\;, \\
\len(\SET{n}(b_0 \ldots b_{n-1},\dstbr{l})) = n\;, \\
\len(\MOV{n}(\srcbr{k},\dstbr{l})) = 3 \mul n\;, \hsp{9.4}
\end{ldispl}%
\begin{ldispl}
\len(\ZPAD{n}{m}(\dstbr{l})) = n - m\;, \\
\len(\MVH{n}{m}(\srcbr{k},\dstbr{l})) = 3 \mul m\;, \\
\len(\MVL{n}{m}(\srcbr{k},\dstbr{l})) = 3 \mul m\;, \\
\len(\MUL{n}(\srcbri{1}{k_1},\srcbri{2}{k_2},\dstbr{l})) =
 36 \mul n^2 + 24 \mul n + 1\;.
\end{ldispl}%

The instruction sequences defined in this section do compute the 
intended operations in case of fully coinciding $n$-bit words.
The instruction sequences defined for addition and transfer of a stored 
word in~\cite{BM13b} do not compute the intended operations in case of 
fully coinciding $n$-bit words. 

\section{Long Multiplication and Karatsuba Multiplication}
\label{sect-KMA}

In this section, we describe and analyze instruction sequences that 
express the long multiplication algorithm and the Karatsuba 
multiplication algorithm, using the definitions given in 
Sections~\ref{sect-words} and~\ref{sect-opns-words}.
The latter algorithm is applicable only if $N \geq 3$.

$\nm{LMUL}_N$ is the instruction sequence described by
\begin{ldispl}
\MUL{N}(I_1,I_2,O) \conc \halt\;.
\end{ldispl}%
We know by Proposition~\ref{prop-MUL-correct} that $\nm{LMUL}_N$ 
computes the function on bit strings that models multiplication on 
natural numbers less than $2^N$ with respect to their binary 
representation by $N$-bit words.
It does so according to the long multiplication algorithm.
\begin{proposition}
\label{prop-LMULi-length}
$\len(\nm{LMUL}_N) = 36 \mul N^2 + 24 \mul N + 2$.
\end{proposition}
\begin{proof}
This is trivial because 
$\len(\nm{LMUL}_N) = \len(\nm{MUL}_N(I_1,I_2,O)) + 1$.
\qed
\end{proof}

$\nm{KMUL}_N$, where $N \geq 3$, is the instruction sequence described 
by
\begin{ldispl}
\MOV{N}(I_1,I_1^{\ceil{\log_2(N - 2)}}) \conc
\MOV{N}(I_2,I_2^{\ceil{\log_2(N - 2)}}) \conc {}
\\
\KMA{N} \conc \MOV{2N}(O^{\ceil{\log_2(N - 2)}},O) \conc \halt \;,
\end{ldispl}%
where $\KMA{n}$ is inductively defined in Table~\ref{table-inseq-KMA}.
\begin{table}[!t]
\caption{\normalsize Definition of $\KMA{n}$ ($1 \leq n \leq N$)}
\label{table-inseq-KMA}
\normalsize
\renewcommand{\arraystretch}{1.544}
$
\begin{array}{@{}l@{}}
\hline
\textrm{if $n \leq 3$ then:}
\\
\KMA{n} = \MUL{n}(\wIi{n},\wIii{n},\dwO{n})\;,
\eqnsep
\textrm{if $n > 3$ then:}
\\
\KMA{n} = {}
\\ \quad
 \MVH{n}{\floor{n/2}}(\wIi{n},\wIi{\floor{n/2}}) \conc
 \MVH{n}{\floor{n/2}}(\wIii{n},\wIii{\floor{n/2}}) \conc {}
\\ \quad
 \KMA{\floor{n/2}} \conc
 \MOV{2 \floor{n/2}}(\dwO{\floor{n/2}},\dwPi{n}) \conc {}
\\ \quad
 \MVL{n}{\ceil{n/2}}(\wIi{n},\wIi{\ceil{n/2}}) \conc
 \MVL{n}{\ceil{n/2}}(\wIii{n},\wIii{\ceil{n/2}}) \conc {}
\\ \quad
 \KMA{\ceil{n/2}} \conc
 \MOV{2 \ceil{n/2}}(\dwO{\ceil{n/2}},\dwPii{n}) \conc {}
\\ \quad
 \MVH{n}{\floor{n/2}}(\wIi{n},\dwTi) \conc
 \ZPAD{\ceil{n/2}+1}{\floor{n/2}}(\dwTi) \conc {}
\\ \quad
 \MVL{n}{\ceil{n/2}}(\wIi{n},\dwTii) \conc
 \ZPAD{\ceil{n/2}+1}{\ceil{n/2}}(\dwTii) \conc
 \ADD{\ceil{n/2}+1}(\dwTi,\dwTii,\wIi{\ceil{n/2}+1}) \conc {}
\\ \quad
 \MVH{n}{\floor{n/2}}(\wIii{n},\dwTi) \conc
 \ZPAD{\ceil{n/2}+1}{\floor{n/2}}(\dwTi) \conc {}
\\ \quad
 \MVL{n}{\ceil{n/2}}(\wIii{n},\dwTii) \conc
 \ZPAD{\ceil{n/2}+1}{\ceil{n/2}}(\dwTii) \conc
 \ADD{\ceil{n/2}+1}(\dwTi,\dwTii,\wIii{\ceil{n/2}+1}) \conc {}
\\ \quad
 \KMA{\ceil{n/2}+1} \conc
 \MOV{2(\ceil{n/2}+1)}(\dwO{\ceil{n/2}+1},\dwPiii{n}) \conc {}
\\ \quad
 \ZPAD{2(\ceil{n/2}+1)}{2 \floor{n/2}}(\dwPi{n}) \conc 
 \ZPAD{2(\ceil{n/2}+1)}{2 \ceil{n/2}}(\dwPii{n}) \conc {}
\\ \quad
 \SUB{2(\ceil{n/2}+1)}(\dwPiii{n},\dwPi{n},\dwTi) \conc
 \SUB{2(\ceil{n/2}+1)}(\dwTi,\dwPii{n},\dwTi) \conc {}
\\ \quad
 \ZPAD{2n}{2(\ceil{n/2}+1)}(\dwPi{n}) \conc 
 \ZPAD{2n}{2(\ceil{n/2}+1)}(\dwPii{n}) \conc 
 \ZPAD{2n}{2(\ceil{n/2}+1)}(\dwTi) \conc {}
\\ \quad
 \SHL{2n}{2\ceil{n/2}}(\dwPi{n},\dwTii) \conc
 \SHL{2n}{\ceil{n/2}}(\dwTi,\dwTi) \conc {}
\\ \quad
 \ADD{2n}(\dwTii,\dwTi,\dwTi) \conc
 \ADD{2n}(\dwTi,\dwPii{n},\dwO{n})\;,
\\
\textrm{where $\ell(m) = \ceil{\log_2(m - 2)}$.}
\\ \hline
\end{array}
$
\end{table}
$\nm{KMUL}_N$ computes the function on bit strings that models 
multiplication on natural numbers less than $2^N$ with respect to their 
binary representation by $n$-bit words according to the Karatsuba 
multiplication algorithm.

In order to compute the binary representation of the product of two 
natural numbers with binary representations of length $n$ by dividing 
the computation into the computations of the binary representations 
of three products as required by the Karatsuba multiplication algorithm, 
the instruction sequence $\KMA{n}$ contains the instruction sequences 
$\KMA{\floor{n/2}}$, $\KMA{\ceil{n/2}}$, and $\KMA{\ceil{n/2} + 1}$.
Each of these three instruction sequences is immediately preceded by an 
instruction sequence that transfers the binary representations of the 
two natural numbers of which it has to compute the binary representation 
of their product into the appropriate Boolean registers for the 
instruction sequence concerned.
Moreover, each of these three instruction sequences is immediately 
followed by an instruction sequence that transfers the binary 
representation of the product that it has computed into the appropriate 
Boolean registers for $\KMA{n}$.
The tail end of $\KMA{n}$ completes the computation by performing some
operations on the three binary representations of products computed 
before as required by the Karatsuba multiplication algorithm.
For the rest, instruction sequences for zero padding are scattered over 
$\KMA{n}$ where necessary to obtain the locally right length of binary 
representations of natural numbers.

\begin{proposition}
\label{prop-KMUL-correct}
If $N \geq 3$, then the function on bit strings of length $N$ computed 
by $\nm{KMUL}_N$ models multiplication on natural numbers less than 
$2^N$ with respect to their binary representation by $N$-bit words.
\end{proposition}
\begin{proof}
It is straightforward to prove this by induction on $N$, using the 
equations from Section~\ref{sect-sketch-KMA} that form the basis of the 
Karatsuba multiplication algorithm and
Propositions~\ref{prop-basic-operations-correct}, 
\ref{prop-transfer-operations-correct}, and~\ref{prop-MUL-correct}.
\qed
\end{proof}

The following proposition gives a lower estimate and an upper estimate 
for the length of $\nm{KMUL}_N$.
\begin{proposition}
\label{prop-KMUL-length}
If $N \geq 3$, then: 
\begin{ldispl}
\len(\nm{KMUL}_N) \geq
1184 \mul 3^{\floor{\log_2(N)}-1} - 716 \mul 2^{\floor{\log_2(N)}-1} + 
12 \mul N - 70\;,
\\
\len(\nm{KMUL}_N) \leq
1005 \mul 3^\ceil{\log_2(N - 2)}  - 358 \mul 2^\ceil{\log_2(N - 2)} + 
12 \mul N - 249\;.
\end{ldispl}%
\end{proposition}
\begin{proof}
Because $\len(\nm{KMUL}_N) = \len(\KMA{N}) + 12 \mul N + 1$, we have to 
prove that
\begin{ldispl}
\len(\KMA{N}) \geq
1184 \mul 3^{\floor{\log_2(N)}-1} - 716 \mul 2^{\floor{\log_2(N)}-1} - 
71\;,
\\
\len(\KMA{N}) \leq 
1005 \mul 3^\ceil{\log_2(N - 2)} - 358 \mul 2^\ceil{\log_2(N - 2)} - 
250\;.
\end{ldispl}%
Let $c_1 = \len(\MUL{1})$, $c_2 = \len(\MUL{2})$, $c_3 = \len(\MUL{3})$, 
and for each $n > 3$, 
$c_n = 
 \len(\KMA{n}) - \len(\KMA{\floor{n/2}}) - \len(\KMA{\ceil{n/2}}) -
 \len(\KMA{\ceil{n/2}+1})$.
Using the already calculated lengths of the parameterized instruction 
sequences defined in Section~\ref{sect-opns-words}, we obtain by simple 
calculations that $c_1 = 61$, $c_2 = 193$, $c_3 = 397$, and for each 
$n > 3$, $c_n = 126 \mul \ceil{n/2} + 116 \mul n + 142$.
Let $c'_0 = c_3$, $c''_0 = c_3$, and for each $m > 0$, 
$c'_m = c_{2^m+2}$ and $c''_m = c_{2^{m+1}}$.
In other words, $c'_0 = 397$, $c''_0 = 397$, and for each $m > 0$, 
$c'_m = 358 \mul 2^{m-1} + 500$ and $c''_m = 358 \mul 2^m + 142$.
Because $\floor{x} = k$ iff $k \leq x < k + 1$, 
$\ceil{x} = k$ iff $k - 1 < x \leq k$, and 
$\log_2(x) = y$ iff $x = 2^y$, it is clear that 
$c_n \leq c'_m$  if $m = \ceil{\log_2(n-2)}$ and
$c_n \geq c''_m$ if $m = \floor{\log_2(n)} - 1$.

Let $M = \ceil{\log_2(N-2)}$, and let $m \leq M$.
It follows directly from the proof of the proposition at the end of 
Section~\ref{sect-words} that, for all $n$ such that 
$m = \ceil{\log_2(n-2)}$, the deepest level of recursion at which 
$\KMA{n}$ occurs is $M - m$.
Moreover, it follows directly from the definition of $\KMA{n}$ that, 
for all $n > 0$, $\KMA{n}$ occurs at this level only if $n$ is less 
than or equal to the greatest $n'$ such that $m = \ceil{\log_2(n'-2)}$.
We also have that $c_n \leq c_{n'}$ if $n \leq n'$, and 
$c_{n'} \leq c'_m$ if $m = \ceil{\log_2(n'-2)}$.
All this means that 
$\len(\KMA{N}) \leq \sum_{i=0}^M (c'_i \mul 3^{M-i})$.
In other words,
$\len(\KMA{N}) \leq 
 397 \mul 3^M + \sum_{i=1}^M ((358 \mul 2^{i-1} + 500) \mul 3^{M-i})$.
Using elementary properties of sums and the property that 
$\sum_{i=0}^k x^i = (1 - x^{k+1}) / (1-x)$, we obtain 
$397 \mul 3^M + \sum_{i=1}^M ((358 \mul 2^{i-1} + 500) \mul 3^{M-i}) =
 397 \mul 3^M + 358 \mul (3^M - 2^M) + \linebreak
 500 \mul ((3^M - 1) / 2) = 
 1005 \mul 3^M - 358 \mul 2^M - 250$.
Hence, because $M = \linebreak \ceil{\log_2(N-2)}$,
$\len(\KMA{N}) \leq 
 1005 \mul 3^\ceil{\log_2(N-2)} - 358 \mul 2^\ceil{\log_2(N-2)} - 250$.

Let $M' = \floor{\log_2(N)} - 1$, and let $m \leq M'$.
We can show similarly to above that, for all $n$ such that 
$m = \floor{\log_2(n)} - 1$, the least deep level of recursion at which 
$\KMA{n}$ occurs is $M' - m$.
Moreover, it follows directly from the definition of $\KMA{n}$ that, 
for all $n > 0$, $\KMA{n}$ occurs at this level only if $n$ is greater 
than or equal to the least $n'$ such that 
$m = \floor{\log_2(n')} - 1$.
We also have that $c_n \geq c_{n'}$ if $n \geq n'$, and 
$c_{n'} \geq c''_m$ if $m = \floor{\log_2(n')} - 1$.
All this means that 
$\len(\KMA{N}) \geq \sum_{i=0}^{M'} (c''_i \mul 3^{M'-i})$.
In other words,
$\len(\KMA{N}) \geq 
 397 \mul 3^{M'} + 
 \sum_{i=1}^{M'} ((358 \mul 2^i + 142) \mul 3^{M'-i})$.
Using the same properties of sums as before, we obtain 
$397 \mul 3^{M'} + 
 \sum_{i=1}^{M'} ((358 \mul 2^i + 142) \mul 3^{M'-i}) =
 397 \mul 3^{M'} + \linebreak 358 \mul (2 \mul (3^{M'} - 2^{M'})) + 
 142 \mul ((3^{M'} - 1) / 2) = 
 1184 \mul 3^{M'} - 716 \mul 2^{M'} - 71$. 
Hence, because $M' = \floor{\log_2(N)} - 1$,
$\len(\KMA{N}) \geq 
 1184 \mul 3^{\floor{\log_2(N)}-1} - \linebreak
 716 \mul 2^{\floor{\log_2(N)}-1} - 71$.
\qed
\end{proof}
It is unclear to us whether it is practically possible to improve the 
lower estimate and upper estimate for the length of $\nm{KMUL}_N$ 
considerably.

The following is a corollary of Propositions~\ref{prop-LMULi-length} 
and~\ref{prop-KMUL-length}.
\begin{corollary}
\label{corollary-LMULi-KMUL-length-1}
$\len(\nm{LMUL}_N) = \Theta(N^2)$ and
$\len(\nm{KMUL}_N) = \Theta(N^{\log_2(3)}) = \Theta(N^{1,5849\ldots})$.
\end{corollary}
This corollary can be paraphrased as follows: the length of the 
instruction sequences $\nm{LMUL}_N$ and $\nm{KMUL}_N$, which express the 
long multiplication algorithm and the Karatsuba multiplication 
algorithm, are asymptotically bounded, up to a constant 
factor, both above and below by $N^2$ and $N^{\log_2(3)}$, respectively.
It is striking because these algorithms are known to compute the 
function that models multiplication on natural numbers less than $2^N$ 
with respect to their binary representation by $N$-bit words also in 
time asymptotically bounded, up to a constant factor, both above and 
below by $N^2$ and $N^{\log_2(3)}$, respectively.
This suggests, like some results from~\cite{BM13a}, that instruction 
sequence size and computation time are polynomially related measures.

Using Propositions~\ref{prop-LMULi-length} and~\ref{prop-KMUL-length}, 
it is easy to check that 
(a)~$\nm{LMUL}_N$ is longer than $\nm{KMUL}_N$ only if  $N > 264$ and
(b)~$\nm{LMUL}_N$ is longer than $\nm{KMUL}_N$ if $N > 6666$.
On that account, the following is another corollary of 
Propositions~\ref{prop-LMULi-length} and~\ref{prop-KMUL-length}.
\begin{corollary}
\label{corollary-LMULi-KMUL-length-2}
$N > 2^8$ if $\len(\nm{LMUL}_N) > \len(\nm{KMUL}_N)$ and
$\len(\nm{LMUL}_N) > \len(\nm{KMUL}_N)$ if $N > 2^{13}$.
\end{corollary}
In the area of algorithm efficiency, like in the area of computational 
complexity, the focus is mainly on asymptotic properties of algorithms, 
like Corollary~\ref{corollary-LMULi-KMUL-length-1}.
To our knowledge, there is virtually no attention in this area to 
properties related to crossover points between algorithms, like
Corollary~\ref{corollary-LMULi-KMUL-length-2}.
We think that properties of the latter kind are frequently more relevant 
to practice than properties of the former kind.
However, existing knowledge about crossover points between algorithms 
is mainly based on experimental data which are highly dependent on the
computer, operating system, programming language and compiler used in 
the experiment. 
Moreover, if this kind of knowledge is referred to at all, it is often 
turned into the form of a rule of thumb.
For example, the following statement and minor variants of it can be 
found at many places (webpages, articles, and books) without further 
justification: ``As a rule of thumb, Karatsuba is usually faster when 
the multiplicands are longer than 320–-640 bits'' 
(see e.g.~\cite{Wik13a}).  

It is obvious that $\nm{LMUL}_N$ and $\nm{KMUL}_N$ need the same number 
of input registers and the same number of output registers. 
However, the number of auxiliary registers used by $\nm{KMUL}_N$ is 
always greater than the number of auxiliary registers used by 
$\nm{LMUL}_N$.
The number of auxiliary registers used by $\nm{KMUL}_N$ is 
$10 \mul N \mul \ceil{\log_2(N - 2)} + 18 \mul N + 1$ 
and the number of auxiliary registers used by $\nm{LMUL}_N$ is only
$4 \mul N + 1$.
In the instance that $N = 2^8$, these numbers correspond to $\pm 3$K 
bytes and $\pm 128$ bytes, respectively; and
in the instance that $N = 2^{13}$, these numbers correspond to 
$\pm 148$K bytes and $\pm 4$K bytes, respectively.

In this paper, we do not answer the question whether there exist 
instruction sequences shorter than $\nm{LMUL}_N$ and $\nm{KMUL}_N$ that 
express the long multiplication algorithm and Karatsuba multiplication 
algorithm, respectively.
The practical problem with proving or disproving the existence of 
shorter instruction sequences is that it needs basically an extremely
extensive case distinction.
We expect that, if the length of $\nm{LMUL}_N$ and/or $\nm{KMUL}_N$ can 
be reduced, it cannot be reduced much. 
The reason for this is that we have striven in 
Section~\ref{sect-opns-words} for instruction sequences without 
unreachable subsequences, different suffixes with the same behaviour on 
execution, and jump instruction that can be eliminated without 
introducing different suffixes with the same behaviour on execution.

\section{Long Multiplication and Backward Jump Instructions}
\label{sect-BJMP}

In this section, a minor variant of the long multiplication algorithm is
expressed by an instruction sequence that contains a backward jump 
instruction in addition to instructions to set and get the content of 
Boolean registers, forward jump instructions, and a termination 
instruction.

We use the fragment without repetition operator of an extension of \PGA\ 
with, for each $l \in \Nat$, a \emph{backward jump instruction} 
$\bjmp{l}$ as additional primitive instruction.
On execution of an instruction sequence, the effect of a backward jump 
instruction $\bjmp{l}$ is that execution proceeds with the $l$th 
previous primitive instruction of the instruction sequence concerned --- 
if $l$ equals $0$ or there is no primitive instruction to proceed with, 
inaction occurs.
We write \PGAbj\ for the above-mentioned extension of \PGA.
For a mathematically precise treatment of \PGAbj\ without repetition 
operator, we refer to the treatment of C, which is a variant of \PGA, 
in~\cite{BP09a}.
The fragment of \PGAbj\ without the repetition operator coincides with 
the fragment of C without backward instructions other than backward jump 
instructions.

The additional basic operations on words that are relevant in this 
section are the operations that model Euclidean division by $2^m$,
decrement by $1$, and nonzero test on natural numbers less than $2^n$, 
with respect to their representation by $n$-bit words ($0 < n \leq N$, 
$0 < m < n$).
The operation modeling Euclidean division by $2^m$ is commonly known as 
``shift right by $m$ positions''.
For these operations, we define parameterized instruction sequences 
computing them in case the parameters are properly instantiated (see 
below):
\begin{ldispl}
\SHR{n}{m}(\srcbr{k},\dstbr{l}) \deq {}
\\ \quad
\Conc{i = 0}{n-1-m} 
 (\ptst{\srcbr{k{+}m{+}i}.\getbr} \conc 
  \ntst{\dstbr{l{+}i}.\setbr{\True}} \conc
  \dstbr{l{+}i}.\setbr{\False}) \conc {}
\\[.5ex] \quad
\Conc{i = 0}{m-1} (\dstbr{l{+}n{-}m{+}i}.\setbr{\False})\;,
\hsp{17.75} 
\end{ldispl}%
\begin{ldispl}
\DEC{n}(\srcbr{k},\dstbr{l}) \deq {}
\\ \quad
\Conc{i = 0}{n-1} 
 (\ntst{\srcbr{k{+}i}.\getbr} \conc \fjmp{3} \conc
  \dstbr{l{+}i}.\setbr{\False} \conc \fjmp{5} \conc
  \dstbr{l{+}i}.\setbr{\True}) \conc
\fjmp{1} \conc \fjmp{1} \conc \fjmp{1}\;,
\eqnsep
\TSTNZ{n}(\srcbr{k}) \deq {}
\\ \quad
\Conc{i = 0}{n-1} 
 (\ptst{\srcbr{k{+}i}.\getbr} \conc \fjmp{2}) \conc \fjmp{2}\;, 
\end{ldispl}%
where 
$s$ ranges over $\set{\mathsf{in},\mathsf{aux}}$, 
$d$ ranges over $\set{\mathsf{aux},\mathsf{out}}$, and
$k,l$ range over $\Natpos$.
For each of the first two parameterized instruction sequences, the first 
parameter correspond to the operand of the operation concerned and the 
second parameter corresponds to the result of the operation concerned.
The intended operations are computed provided that the instantiation of 
the first parameter and the instantiation of the second parameters do 
not lead to partially coinciding $n$-bit words.
In this section, this condition will always be satisfied.
No result is stored on execution of $\TSTNZ{n}$.
Instead, the first primitive instruction following $\TSTNZ{n}$ is 
skipped if the nonzero test fails.

\begin{proposition}
\label{prop-add-basic-operations-correct}
Let $n,m \in \Nat$ be such that $0 < n \leq N$ and $0 < m < n$.
Then the function on bit strings of length $n$ computed by 
\begin{enumerate}
\item
$\SHR{n}{m}(I_1,O) \conc \halt$ models Euclidean division by $2^m$ 
modulo $2^n$ on natural numbers less than $2^n$ with respect to their 
binary representation by $n$-bit words;
\item
$\DEC{n}(I_1,O) \conc \halt$ models subtraction by $1$ modulo $2^n$ on 
natural numbers less than $2^n$ with respect to their binary 
representation by $n$-bit words;
\item
$\TSTNZ{n}(I_1) \conc
 \ptst{O.\setbr{\True}} \conc O.\setbr{\False} \conc \halt$ 
models the function $\nm{isnz}$ from natural num\-bers less than $2^n$ 
to natural numbers less than $2^1$ defined by $\nm{isnz}(0) = 0$ and 
$\nm{isnz}(k + 1) = 1$ with respect to their binary representation by 
$n$-bit words and $1$-bit words, respectively.
\end{enumerate}
\end{proposition}
\begin{proof}
Each of these properties is easy to prove by induction on $n$ with case 
distinction on the content of the input register containing the most 
significant bit of the operand of the operation concerned in both the 
basis step and the inductive step.
\qed
\end{proof}

The lengths of the parameterized instruction sequences defined above are 
as follows:
\begin{ldispl}
\len(\SHR{n}{m}(\srcbr{k},\dstbr{l})) = 3 \mul n - 2 \mul m\;, \\
\len(\DEC{n}(\srcbr{k},\dstbr{l})) = 5 \mul n + 3\;, \\
\len(\TSTNZ{n}(\srcbr{k})) = 2 \mul n + 1\;.
\end{ldispl}%

For each bit of the representation of the multiplier, $\LMULi{N}$ 
contains a different instruction sequence.
This seems to exclude the use of backward jump instructions to obtain 
an instruction sequence of significantly shorter length, unless 
provision is made for some form of indirect addressing for Boolean 
registers.
However, there exists a minor variant of the long multiplication 
algorithm that makes it possible to have the same instruction sequence 
for each bit of the representation of the multiplier.
From the least significant bit of the representation of the multiplier
onwards, the algorithm concerned shifts the representation of the 
multiplier by one position to the right after it has dealt with a bit.
In this way, the next bit remains the least significant one throughout.

We proceed with describing an instruction sequence without backward jump 
instructions that expresses this minor variant of the long multiplication 
algorithm.

$\LMULii{N}$ is the instruction sequence described by 
\begin{ldispl}
\MOV{N}(I_1,S_1) \conc \ZPAD{2N}{N}(S_1) \conc
\SET{2N}(0^{2N},S_2) \conc \MOV{N}(I_2,T_1) \conc {}
\\ 
 \bigl(
  \ntst{T_1.\getbr} \conc \fjmp{l} \conc \ADD{2N}(S_1,S_2,S_2) \conc
  \SHL{2N}{1}(S_1,S_1) \conc \SHR{N}{1}(T_1,T_1)
 \bigr)^N \conc {}
\\ 
\MOV{2N}(S_2,O) \conc \halt\;, 
\end{ldispl}%
\begin{ldispl}
\mbox{where} \hsp{28.05} 
\\[1.1ex]
l = \len(\ADD{2N}(S_1,S_2,S_2)) + 1 = 42 \mul N + 2\;.
\end{ldispl}%
 
\begin{proposition}
\label{prop-LMULii-correct}
The function on bit strings of length $N$ computed by $\LMULii{N}$ 
models multiplication on natural numbers less than $2^N$ with respect to 
their binary representation by $N$-bit words.
\end{proposition}
\begin{proof}
We prove a stronger property that also covers the final contents of the 
$2N$ successive auxiliary registers starting with the one named $S_1$, 
the $2N$ successive auxiliary registers starting with the one named 
$S_2$, and the $N$ successive auxiliary registers starting with the one 
named $T_1$.
This stronger property is straightforward to prove, using 
Propositions~\ref{prop-basic-operations-correct}, 
\ref{prop-transfer-operations-correct}, 
and~\ref{prop-add-basic-operations-correct},
by induction on $N$ with case distinction on the content of the input 
register containing the most significant bit of the second operand of 
the operation concerned in both the basis step and the inductive step.
\qed
\end{proof}

\begin{proposition}
\label{prop-LMULii-length}
$\len(\LMULii{N}) = 51 \mul N^2 + 14 \mul N + 1$.
\end{proposition}
\begin{proof}
This is a matter of simple additions, subtractions, and multiplications.
\linebreak[2]
\qed
\end{proof}

The following is a corollary of Propositions~\ref{prop-LMULi-length} 
and~\ref{prop-LMULii-length}.
\begin{corollary}
$\len(\LMULii{N}) > \len(\LMULi{N})$.
\end{corollary}

For each bit of the representation of the multiplier, $\LMULii{N}$ 
contains the same instruction sequence.
That is, it contains $N$ duplicates of the same instruction sequence.
This duplication can be eliminated by implementing a for loop by means
of a backward jump instruction.

We proceed with describing an instruction sequence with a backward jump 
instruction that expresses the minor variant of the long multiplication 
algorithm.
We write $\underline{N}$ for the shortest representation of the natural 
number $N$ in the binary number system.

$\LMULiii{N}$ is the instruction sequence described by 
\begin{ldispl}
\MOV{N}(I_1,S_1) \conc \ZPAD{2N}{N}(S_1) \conc
\SET{2N}(0^{2N},S_2) \conc \MOV{N}(I_2,T_1) \conc {}
\\ 
\SET{\floor{\log_2(N)}+1}(\underline{N},T_2) \conc {}
\\ 
\ntst{T_1.\getbr} \conc \fjmp{l_1} \conc \ADD{2N}(S_1,S_2,S_2) \conc
\SHL{2N}{1}(S_1,S_1) \conc \SHR{N}{1}(T_1,T_1) \conc {}
\\ 
\DEC{\floor{\log_2(N)}+1}(T_2,T_2) \conc
\TSTNZ{\floor{\log_2(N)}+1}(T_2) \conc \bjmp{l_2} \conc {}
\\
\MOV{2N}(S_2,O) \conc \halt\;,
\\[1.1ex]
\mbox{where} 
\\[1.1ex]
l_1 = \len(\ADD{2N}(S_1,S_2,S_2)) + 1 = 42 \mul N + 2\;,
\\
l_2 = \len(\ntst{T_1.\getbr} \conc \ldots \conc 
           \TSTNZ{\floor{\log_2(N)}+1}(T_2)) = 
       51 \mul N + 7 \mul \floor{\log_2(N)} + 10\;.
\end{ldispl}%

\begin{proposition}
\label{prop-LMULiii-correct}
The function on bit strings of length $N$ computed by $\LMULiii{N}$ 
models multiplication on natural numbers less than $2^N$ with respect to 
their binary representation by $N$-bit words.
\end{proposition}
\begin{proof}
We prove a stronger property that also covers the final contents of the 
$2N$ successive auxiliary registers starting with the one named $S_1$, 
the $2N$ successive auxiliary registers starting with the one named 
$S_2$, the $N$ successive auxiliary registers starting with the one 
named $T_1$, and the $\floor{\log_2(N)}+1$ successive auxiliary 
registers starting with the one named $T_2$.
This stronger property is straightforward to prove, using 
Propositions~\ref{prop-basic-operations-correct}, 
\ref{prop-transfer-operations-correct}, 
and~\ref{prop-add-basic-operations-correct},
by induction on $N$ with case distinction on the content of the input 
register containing the most significant bit of the second operand of 
the operation concerned in both the basis step and the inductive step.
\qed
\end{proof}

\begin{proposition}
\label{prop-LMULiii-length}
$\len(\LMULiii{N}) = 66 \mul N + 8 \mul \floor{\log_2(N)} + 13$.
\end{proposition}
\begin{proof}
This is a matter of simple additions, subtractions, and multiplications.
\linebreak[2]
\qed
\end{proof}

The following is a corollary of Propositions~\ref{prop-LMULi-length}, 
\ref{prop-LMULii-length}, and~\ref{prop-LMULiii-length}.
\begin{corollary}
$\len(\LMULiii{N}) = \Theta(N)$ while both
$\len(\LMULi{N}) = \Theta(N^2)$, and $\len(\LMULii{N}) = \Theta(N^2)$.
\end{corollary}
Hence, $\LMULiii{N}$ is asymptotically shorter than both $\LMULi{N}$ and
$\LMULii{N}$.
By Corollary~\ref{corollary-LMULi-KMUL-length-1}, we know that 
$\LMULiii{N}$ is asymptotically shorter than $\nm{KMUL}_N$ too.

The following is a corollary of Propositions~\ref{prop-LMULi-length}, 
\ref{prop-KMUL-length}, \ref{prop-LMULii-length}, 
and~\ref{prop-LMULiii-length}.
\begin{corollary}
$\len(\LMULiii{N}) < \len(\LMULi{N})$ and
$\len(\LMULiii{N}) < \len(\LMULii{N})$ if $N > 1$, and what is more, 
$\len(\LMULiii{N}) < \len(\nm{KMUL}_N)$ if $N > 2$.
\end{corollary}
Hence, $\LMULiii{N}$ is already shorter than $\LMULi{N}$, $\LMULii{N}$, 
and $\nm{KMUL}_N$ if $N$ is still very small.
In fact, long multiplication is non-trivial only if $N > 1$ and 
Karatsuba multiplication is applicable only if $N > 2$.

\section{Long Multiplication and the Halting Problem}
\label{sect-HP}

In this section, we argue that the instruction sequences $\LMULii{N}$ 
and $\LMULiii{N}$ from Section~\ref{sect-BJMP} form a hard witness of 
the inevitable existence of a halting problem in the practice of 
imperative programming.

Turing's result regarding the undecidability of the halting problem 
(see e.g.~\cite{Tur37a}) is a result about Turing machines.
In~\cite{BM09k}, we consider it as a result about programs rather than 
machines, taking instruction sequences as programs.
The instruction sequences concerned are essentially the finite 
instruction sequences that can be denoted by closed \PGAbj\ terms.
Unlike in the current paper, the basic instructions are not fixed, but 
their effects are restricted to the manipulation of something that can 
be understood as the content of the tape of a Turing machine with a 
specific tape alphabet, together with the position of the tape head.
Different choices of basic instructions give rise to different halting 
problem instances and one of these instances is essentially the same as 
the halting problem for Turing machines.
Because of their orientation to Turing machines, we consider all 
instances treated in~\cite{BM09k} theoretical halting problem instances.

All halting problem instances would evaporate if the instruction 
sequences concerned would be restricted to the ones without backward 
jump instructions.
This is irrespective of whether the effects of the basic instructions 
have anything to do with the manipulation of a Turing machine tape.
In the case that we have basic instructions to set and get the content 
of Boolean registers, instruction sequences without backward jump 
instructions are sufficient to compute all functions 
$\funct{f}{\set{0,1}^n}{\set{0,1}^m}$ ($n,m \in \Nat$).
This raises the question whether there exists a good reason for not 
abandoning backward jump instructions altogether in such cases.
The function that models multiplication on natural numbers less than 
$2^N$ with respect to their binary representation by $N$-bit words 
offers a good reason: the length of the instruction sequence that 
computes it according to the long multiplication algorithm can be 
reduced significantly by the use of backward jump instructions.
The length of the instruction sequence that computes this function can 
be reduced even more by the use of backward jump instructions than by 
going over to one of the multiplication algorithms that are known to 
yield shorter instruction sequences without backward jump instructions 
than the long multiplication algorithm such as for example the Karatsuba 
multiplication algorithm.

Thus, the instruction sequences $\LMULii{N}$ and $\LMULiii{N}$ form a 
hard witness of the inevitable existence of a halting problem in the 
practice of imperative programming, where programs must have manageable 
size.
Because of its orientation to actual programming, we consider the 
halting problem for the instruction sequences with forward and backward 
jump instructions, and with only basic instructions to set and get the 
content of Boolean registers, a practical halting problem.
It is unknown to us whether there is a connection between the 
solvability or unsolvability of the halting problem for these 
instruction sequences and some form of diagonal argument.
It is easy to prove that this halting problem is both NP-hard and 
coNP-hard.
We do not know whether stronger lower bounds for its complexity can be 
found in the literature.
An extensive search for such lower bounds and other results concerning 
this halting problem or a similar halting problem has been unsuccessful.

\section{Concluding Remarks}
\label{sect-concl}

We have described finite instruction sequences, containing only 
instructions to set and get the content of Boolean registers, forward 
jump instructions, and a termination instruction, that compute the 
function that models multiplication on natural numbers less than 
$2^N$ with respect to their binary representation by $N$-bit words
according to the long multiplication algorithm and the Karatsuba 
multiplication algorithm.
We have described those instruction sequences by means of terms of \PGA,
an algebraic theory of single-pass instruction sequences.

Thus, we have provided mathematically precise alternatives to the
natural language and pseudo code descriptions of these multiplication 
algorithms found in mathematics and computer science literature on 
multiplication algorithms.
Moreover, we have calculated the exact size of the instruction sequence 
$\nm{LMUL}_N$ expressing the long multiplication algorithm and lower and 
upper estimates for the size of the instruction sequence $\nm{KMUL}_N$ 
expressing the Karatsuba multiplication algorithm.
We have among other things found that: 
(a)~$\len(\nm{LMUL}_N) = \Theta(N^2)$ and
    $\len(\nm{KMUL}_N) = \Theta(N^{\log_2(3)})$; 
(b)~$N > 2^8$ if $\len(\nm{LMUL}_N) > \len(\nm{KMUL}_N)$, and
    $\len(\nm{LMUL}_N) > \len(\nm{KMUL}_N)$ if $N > 2^{13}$.
It is suggested by~(a) that instruction sequence size and computation 
time are polynomially related measures.
It is still an open question whether this is the case.

As a bonus, we have found that the number of auxiliary registers used by 
$\nm{LMUL}_N$ is $4 \mul N + 1$ and the number of auxiliary registers 
used by $\nm{KMUL}_N$ is 
$10 \mul N \mul \ceil{\log_2(N - 2)} + 18 \mul N + 1$.
It is also an open question whether the number of auxiliary registers 
that are used by an instruction sequence and computation space are 
related measures.

We have also gone into the use of an instruction sequence with backward 
jump instructions for expressing the long multiplication algorithm.
We have described a finite instruction sequence $\LMULiii{N}$ containing 
a backward jump instruction, in addition to the instructions to set and 
get the content of Boolean registers, forward jump instructions, and a 
termination instruction, that expresses a minor variant of the long 
multiplication algorithm.
We have calculated the exact size of this instruction sequence and have 
among other things found that:
(a)~$\len(\LMULiii{N}) = \Theta(N)$;
(b)~$\len(\LMULiii{N}) < \len(\LMULi{N})$ if $N > 1$, and 
    $\len(\LMULiii{N}) < \len(\nm{KMUL}_N)$ if $N > 2$.
Furthermore, we have related these findings to the halting problem.

\subsection*{Acknowledgements}
We thank Dimitri Hendriks from the VU University Amsterdam for carefully
reading a draft of this paper and for pointing out an error in it.

\bibliographystyle{splncs03}
\bibliography{IS}

\end{document}